%% file: QIP_2019_QCBM.tex
\documentclass[%
 preprint,
groupedaddress,
 amsmath,amssymb,
 aps,
floatfix
]{revtex4-1}
\usepackage{amssymb}
\usepackage{amsthm}
\usepackage{amsmath}
\usepackage{etoolbox}
\usepackage{ulem,url}
\usepackage{graphicx}
\usepackage{dcolumn}
\usepackage{bm}
\usepackage{hyperref}
\usepackage[mathlines]{lineno}
\usepackage{cleveref}
\usepackage{color}

\newcommand{\Fscore}{\ensuremath{\mathrm{F}_1} }
\newcommand{\qBAS}{\ensuremath{\mathrm{qBAS22}}}
\newcommand{\KL}[2]{\ensuremath{\mathrm{D}(#1|#2)}}
\begin{document}

\preprint{APS/123-QED}

 \title{Generative model benchmarks for superconducting qubits}

\thanks{This manuscript has been authored by UT-Battelle, LLC, under Contract No. DE-AC0500OR22725 with the U.S. Department of Energy. The United States Government retains and the publisher, by accepting the article for publication, acknowledges that the United States Government retains a non-exclusive, paid-up, irrevocable, world-wide license to publish or reproduce the published form of this manuscript, or allow others to do so, for the United States Government purposes. The Department of Energy will provide public access to these results of federally sponsored research in accordance with the DOE Public Access Plan.}%

\author{Kathleen E. Hamilton}
\author{Eugene F. Dumitrescu}
\author{Raphael C. Pooser}
\affiliation{Computer Science and Engineering Division, Oak Ridge National Laboratory, One Bethel Valley Road, Oak Ridge, TN 37831 USA}


\begin{abstract}
In this work we experimentally demonstrate how generative model training can be used as a benchmark for small ($<5$ qubits) quantum devices.  Performance is quantified using three data analytic metrics: the Kullbeck-Leiber divergence, and two adaptations of the \Fscore score. Using the $2 \times 2$ Bars and Stripes dataset, we train several different circuit constructions for generative modeling with superconducting qubits.  By taking hardware connectivity constraints into consideration, we show that sparsely connected shallow circuits out-perform denser counterparts on noisy hardware.

\end{abstract}

\pacs{Valid PACS appear here}
\maketitle

\section{Introduction}
\label{sec:introduction}
\input{introduction.tex}

\section{Quantum Circuit Born Machines}
\label{sec:theory}
\input{theory_background.tex}

\section{Results}
\label{sec:experiment1}
\input{experiment1_evaluation.tex}

\subsection{QCBM training with noisy qubits}
\label{sec:experiment2}
\input{experiment2_learning.tex}

\section{Discussion}
\label{sec:discussion}
\input{discussion.tex}

\section{Conclusions}
\label{sec:conclusions}
\input{conclusions.tex}

\section{Acknowledgements}
This work was supported as part of the ASCR Testbed Pathfinder Program at Oak Ridge National Laboratory under FWP \#ERKJ332.  This research used quantum computing system resources of the Oak Ridge Leadership Computing Facility, which is a DOE Office of Science User Facility supported under Contract DE-AC05-00OR22725. Oak Ridge National Laboratory manages access to the IBM Q System as part of the IBM Q Network.

The code used to train QCBM circuits on IBM hardware was adapted from open-source software which is publicly available at \url{https://github.com/GiggleLiu/QuantumCircuitBornMachine} courtesy of Jin-Guo Liu and Lei Wang.
\appendix
\input{appendix1.tex}

\bibliographystyle{unsrt}




\end{document}

%% file: introduction.tex
The increasing diversity of programmable noisy intermediate-scale quantum (NISQ) devices has exposed the need for a unified set of benchmark tasks which assess application-centric device capabilities.  Quantum machine learning (QML) has been presented as a useful tool for benchmarking quantum hardware~\cite{Biamonte2017}. Generative model training was recently proposed as a benchmark task~\cite{perdomo2018opportunities,liu2018differentiable,benedetti2018generative} for NISQ devices.  In this work we use non-adversarial training of a generative model to benchmark superconducting qubit devices.  This approach to generative modeling requires training of a single quantum circuit, making it more practical for implementation on current devices.

Generative models, such as adversarial networks~\cite{goodfellow2014generative}, have recently spurred significant interest in the development of quantum circuit analogues~\cite{lloyd2018quantum,dallaire2018quantum} and adversarial quantum circuits training ~\cite{benedetti2018adversarial,hu2018quantum,zeng2018learning}.  The quantum-circuit Born machine (QCBM) is a generative model constructed as a quantum circuit \cite{benedetti2018generative,cheng2018information,liu2018differentiable}. 
Numerical simulation of QCBMs, constructed using the hardware efficient circuit ansatz \cite{kandala2017hardware} with many ($>10$) entangling layers and trained with non-adversarial methods, using data-driven quantum circuit learning (DDQCL), introduced in \cite{benedetti2018generative} can reproduce several classes of discrete and continuous distributions \cite{liu2018differentiable}.  Here we utilize the gradient-based DDQCL methods of \cite{liu2018differentiable}.

In contrast, NISQ devices accumulate errors due to imperfect gates and environmental decoherence effects.  As such, we expect that the depth of useful NISQ circuits to be limited. After this point, the output becomes random as dictated by the noise.  QML-based benchmarking is a practical method to establish the maximal circuit depth. To experimentally test this hypothesis we train a set of shallow circuits ($<3$ entangling layers) which are deployed on IBM's Toyko chip which has $20$ superconducting qubits.  The entangling layers of all circuits considered can be embedded in a two-rung ladder geometry (e.g. IBM's Melbourne chip \cite{IBM_Melbourne}) ensuring portability of our benchmark.  

Guidelines for benchmarking digital QML algorithms have been proposed~\cite{michielsen2017benchmarking} in terms of the output correctness.  For generative models, correctness refers to the model's ability to reproduce the target distribution. Performance is therefore naturally captured by statistical measures describing the similarity of two distributions, such as the Kullback-Leibler divergence and the \Fscore score.

We evaluated several QCBM circuits on superconducting qubits accessed through the IBM Quantum Hub cloud interface.  The QCBM circuit, training methodology, and performance metrics are described in \Cref{sec:theory}. In \Cref{sec:experiment1} we discuss the interplay between circuit design and QCBM performance. Noisy qubits are introducted into QCBM training in \Cref{sec:experiment2}.  While previous experimental results for machine-learning based benchmarks were executed on direct-access ion trap hardware which can implement all-to-all connectivity~\cite{benedetti2018generative}, our results show comparable performance in superconducting qubits as measured by the Kullback-Leiber divergence.

%% file: theory_background.tex
\begin{figure}[htbp]
\centering\includegraphics[width=\linewidth]{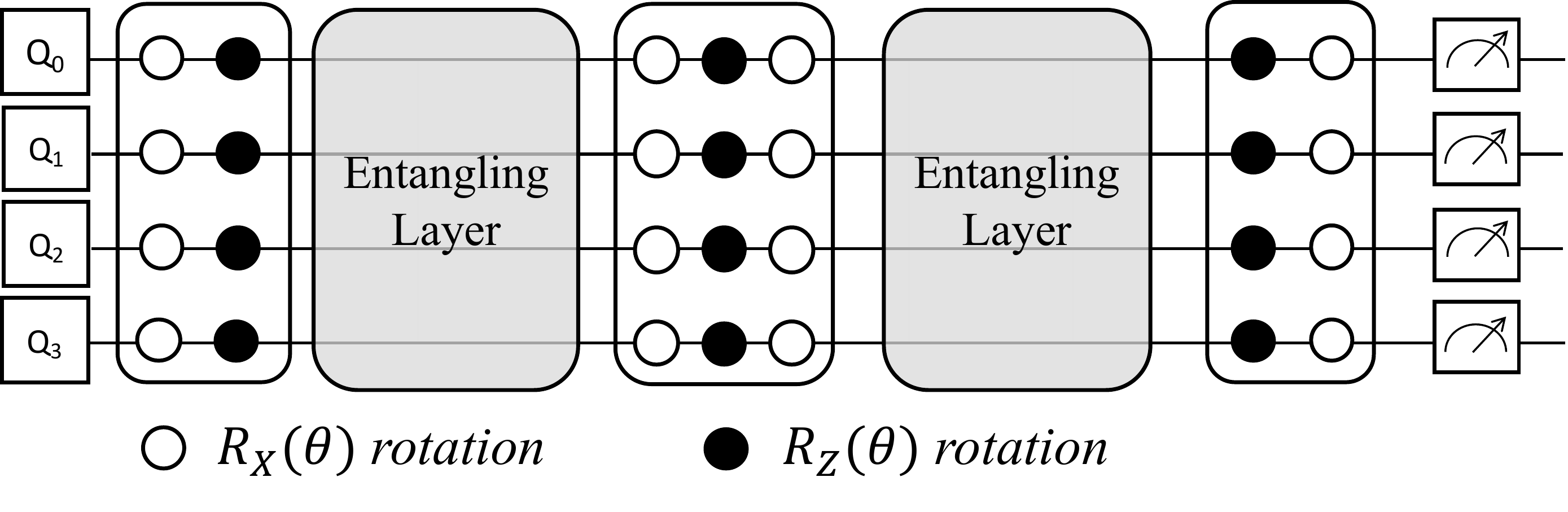}
\caption{The general circuit construction of a QCBM introduced in \cite{benedetti2018generative} is based on the hardware efficient variational quantum eigensolver ansatz of \cite{kandala2017hardware}.}
\label{fig:general_circuit}
\end{figure}

A parametrized quantum circuit defining a particular variational manifold of quantum states is referred to as an \textit{ansatz}.  In this work, as in \cite{liu2018differentiable}, QCBM training is performed with circuits inspired by the hardware efficient ansatz originally applied in the context of the variational quantum eigensolver algorithm~\cite{kandala2017hardware} (see \Cref{fig:general_circuit}).
The BAS(2,2) dataset contains six $2\times2$-pixel black and white striped images. Each image is represented in the computational basis of a $4$-qubit register by fixing a qubit-pixel mapping, and associating black (white) pixels with the states $|0\rangle (|1\rangle)$ (see \Cref{sec:appendix_4}).

While the entangling design introduced in~\cite{liu2018differentiable} contains enough complexity to represent the dataset, for larger image sizes it can require a high degree of qubit connectivity that is not available on current superconducting devices.

\begin{figure}[htbp]
\centering
\includegraphics[width=\linewidth]{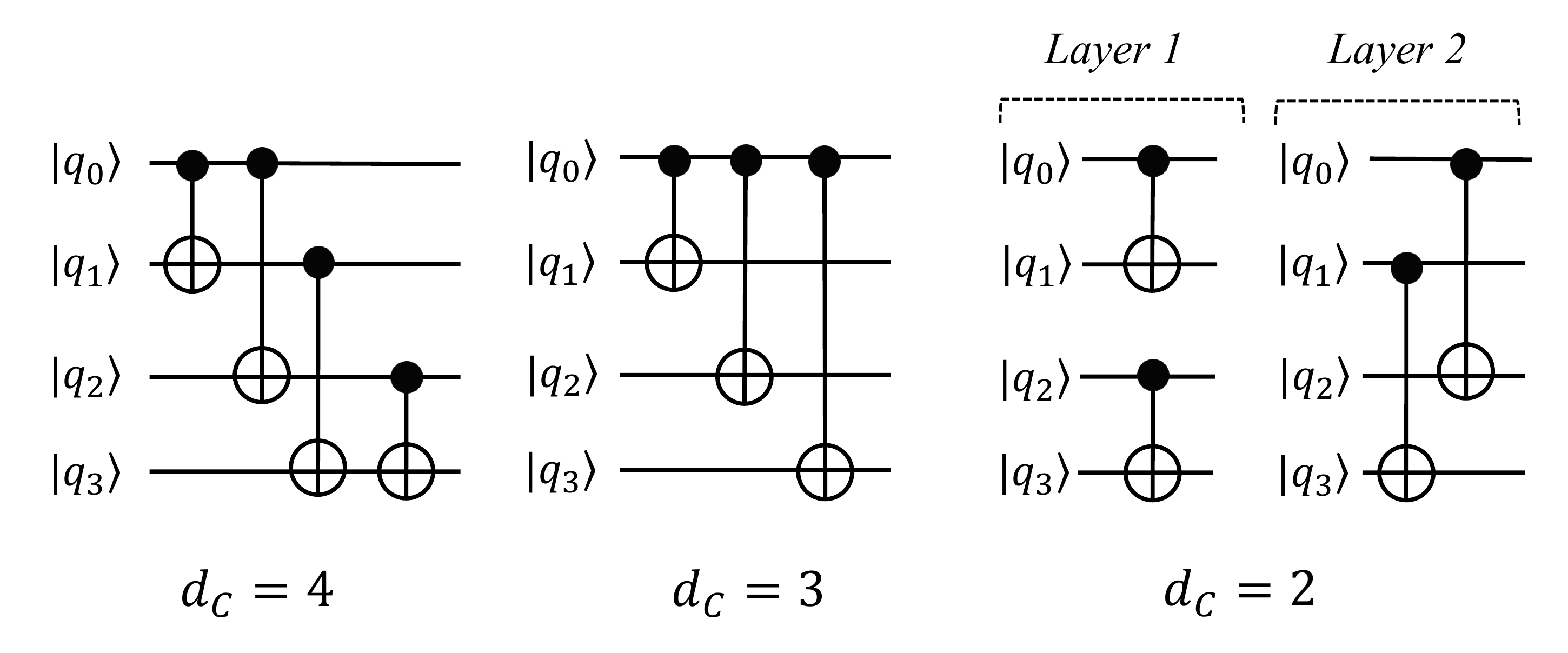}
\caption{The CNOT gate sets used to define individual entangling layers. The $d_C=3$ entangling is the Chow-Liu tree-based design introduced in \cite{liu2018differentiable}.}
\label{fig:entangler_layers}
\end{figure}

To generate BAS(2,2) we train three different ansatz (shown in~\Cref{fig:general_circuit}) whose entangling layers are illustrated in~\Cref{fig:entangler_layers}.  Each circuit is defined on a $4$ qubit register and specified by the number of entangling layers ($L$) and the number of CNOT gates contained within each entangling layer ($d_{C}$).  Current hardware's fixed connectivity presents a challenge when mapping arbitrary datasets. The $d_C=2$ and $d_C=4$ entangling layers conform to IBM's layout, i.e. by restricting CNOT gates to the edges of a $4$ site square plaquette. The $d_C=2$ layers are a sparser circuit construction and only take $\sim 200$ns to apply. 
As CNOTs within a single plaquette cannot be simultaneously applied, we decompose the $d_C=4$ layer into $2$ separate plaquette edge coverings. Thus the $d_C=4$ circuit takes $\sim 400$ ns to apply, adding additional decoherence compared to $d_C=2$. Additionally, since plaquettes may be covered in two ways as shown in~\Cref{fig:entangler_layers}, alternating the two patterns results in a heterogeneous entangling layers structure for $d_C = 2$.  For reference we also use the Chow-Liu tree-based design of \cite{liu2018differentiable} to define circuits with $d_C=3$, though this entangling layer is not embeddable in a single square plaquette.

Many methods exist for training implicit generative models \cite{mohamed2016learning}. In this work the rotational parameters are optimized using Adam \cite{kingma2014Adam}. Overall we follow the training methods described in \cite{liu2018differentiable}: relying on the maximum mean discrepancy (MMD) \cite{gretton2007kernel} to define a loss function for circuit training and using the same unbiased estimator to evaluate the gradient.  In this work we are modeling a known target distribution: we know it is uniform, we can identify the binary states contained in the distribution, and we may sample classically from the distribution without error.

The target distribution $p(x)$ is fixed and defined by the BAS(2,2) dataset. For a given set of rotational parameters we execute a given QCBM circuit, draw $N_{shots}$ samples and label this distribution $q(x)$. 
To compare $q(x)$ to $p(x)$ and quantify the overall QCBM performance we rely on the Kullback-Leiber
(KL) divergence.  The KL divergence compares the two sampled distributions $p(x), q(x)$ by computing the density ratio $p(x_i)/q(x_i)$ of individual states,
\begin{equation}
    \KL{p}{q}=\sum_i p(x_i) \log{\left(\frac{p(x_i)}{q(x_i)}\right)}.
\end{equation}
As $p(x_i)/q(x_i) \to 1$, $\KL{p}{q} \to 0$, but $\KL{p}{q}$ diverges if $p(x_i) \neq 0$ and $q(x_i)=0$.

In addition, the performance metric known as the \Fscore score \cite{goodfellow2016deep} can be used.  We modify the \Fscore score to define an individual value assigned to each BAS(n,m) state and treat the dataset as a $2^{m}+2^{n}-2$ class system.  This metric is analogous to measuring the fidelity of each state and we use it to gain insight into how well each circuit ansatz can learn the states of the BAS(n,m) system. The metric is complementary to $\KL{p}{q}$, giving insight into which eigenstates of the distribution are responsible for high KL values. Further details are given in \Cref{sec:appendix_1}.

We note that the number of samples drawn from a circuit during training can be different from the number of samples taken when evaluating performance metrics.  When evaluating the KL divergence we keep the number of shots fixed at $N_{shots}=2048$, when evaluating the $\qBAS \:$ score (see \Cref{sec:appendix_1}) we keep the number of shots fixed at $N_{shots}=64$.

%% file: experiment1_evaluation.tex
We first use numerical simulation to train each QCBM in order to estimate how well the target distribution can be learned in the absence of noise.
Circuits were constructed using the entangling layers shown in~\Cref{fig:entangler_layers} and trained using the QASM simulator available in IBM Qiskit-Terra.  We limit the number of entangling layers to $L=2$, for a total of $6$ circuits.  Each circuit is trained for $100$ steps of Adam with learning rate $\alpha = 0.2$ and decay rates ($\beta_1 = 0.9, \beta_2=0.999$).  The MMD loss function is calculated using Gaussian kernels with $\sigma = 0.1$.  \Cref{fig:KL_sim_comparision} shows the overall performance of the $3$ circuit ansatz with noiseless qubits for $L=1,2$ when $N_{shots}=1024$ shots are drawn during training.  For each set of rotational parameters, we evaluate the KL divergence of a given circuit $10$ times at every training step with $N_{shots}=2048$ and report the arithmetic mean value of \KL{p}{q}.

\subsection{QCBM training with noiseless qubits}
\label{sec:noiseless_training}
For each value of $\{d_C, L, N_{shots}\}$ a circuit was trained from a random intialization for $\lbrace \theta^{(t=0)} \rbrace$. ~\Cref{tab:sim_L1_values,tab:sim_L2_values} show that for most circuits the shot size used during training has a modest effect on performance for the same circuit (fixed $d_C, L$), however different shot sizes will lead to different trajectories through $\lbrace \theta\rbrace$-space during training.  In particular, the large discrepancy for $d_C=3, L=2$ between $N_{shots}=512$ and $N_{shots}=2048$ is most likely due to $\lbrace \theta\rbrace$ getting trapped in a sub-optimal minimum.  For context, we also trained a non-entangling $L=0,d_C=0$ circuit. With $N_{shots}=1024$ this circuit reached a minimum value of $\KL{p}{q}=1.0(1)$.

\begin{table}[htbp]
\setlength{\tabcolsep}{7pt}
\centering
\caption{$\min(\langle \KL{p}{q}\rangle)$ for $L=1$ circuits simulated on noiseless qubits. Mean calculated over $10$ independent metric evaluations.}
\begin{tabular}{l c c c c}
\hline
L & $N_{shots}$ & $d_C=2$ & $d_C=3$ & $d_C=4$\\
\hline
$1$ & $512$ & $0.95 \pm 0.05$ & $0.33\pm 0.02 $ & $0.24 \pm 0.01$\\
$1$ & $1024$ & $0.93 \pm 0.03$ & $0.34\pm 0.01$ & $0.23\pm 0.01$\\
$1$ & $2048$ & $0.93 \pm 0.03$ & $0.33\pm 0.01$ & $0.23\pm 0.01$\\
\hline
\end{tabular}
\label{tab:sim_L1_values}
\end{table}

\begin{table}[htbp]
\setlength{\tabcolsep}{7pt}
\centering
\caption{$\min(\langle \KL{p}{q}\rangle)$ for $L=2$ circuits simulated on noiseless qubits.  Mean calculated over $10$ independent metric evaluations.}
\begin{tabular}{l c c c c}
\hline
L & $N_{shots}$ & $d_C=2$ & $d_C=3$ & $d_C=4$\\
\hline
 2 & 512 & $0.013\pm 0.004$ & $0.06\pm 0.01$ & $0.02\pm 0.01$\\
 2 & 1024 & $0.088\pm 0.008$ & $0.01\pm 0.01$ & $0.02\pm 0.01$\\
 2 & 2048 & $0.011\pm 0.003$ & $0.13\pm 0.01$ & $0.01\pm 0.01$\\
\hline
\end{tabular}
\label{tab:sim_L2_values}
\end{table}

In general,~\Cref{tab:sim_L1_values,tab:sim_L2_values} show that increasing the complexity of a circuit by increasing the number of rotational parameters will improve performance.  For example, the $d_C=2, L=2$ ($28$ rotational parameters) and $d_C=4,L=1$ ($16$ rotational parameters) circuits contain the same set of CNOT gates, however the better performance is measured with the $d_C=2,L=2$ circuit.

\begin{figure}[htbp]
\centering
\includegraphics[width=\linewidth]{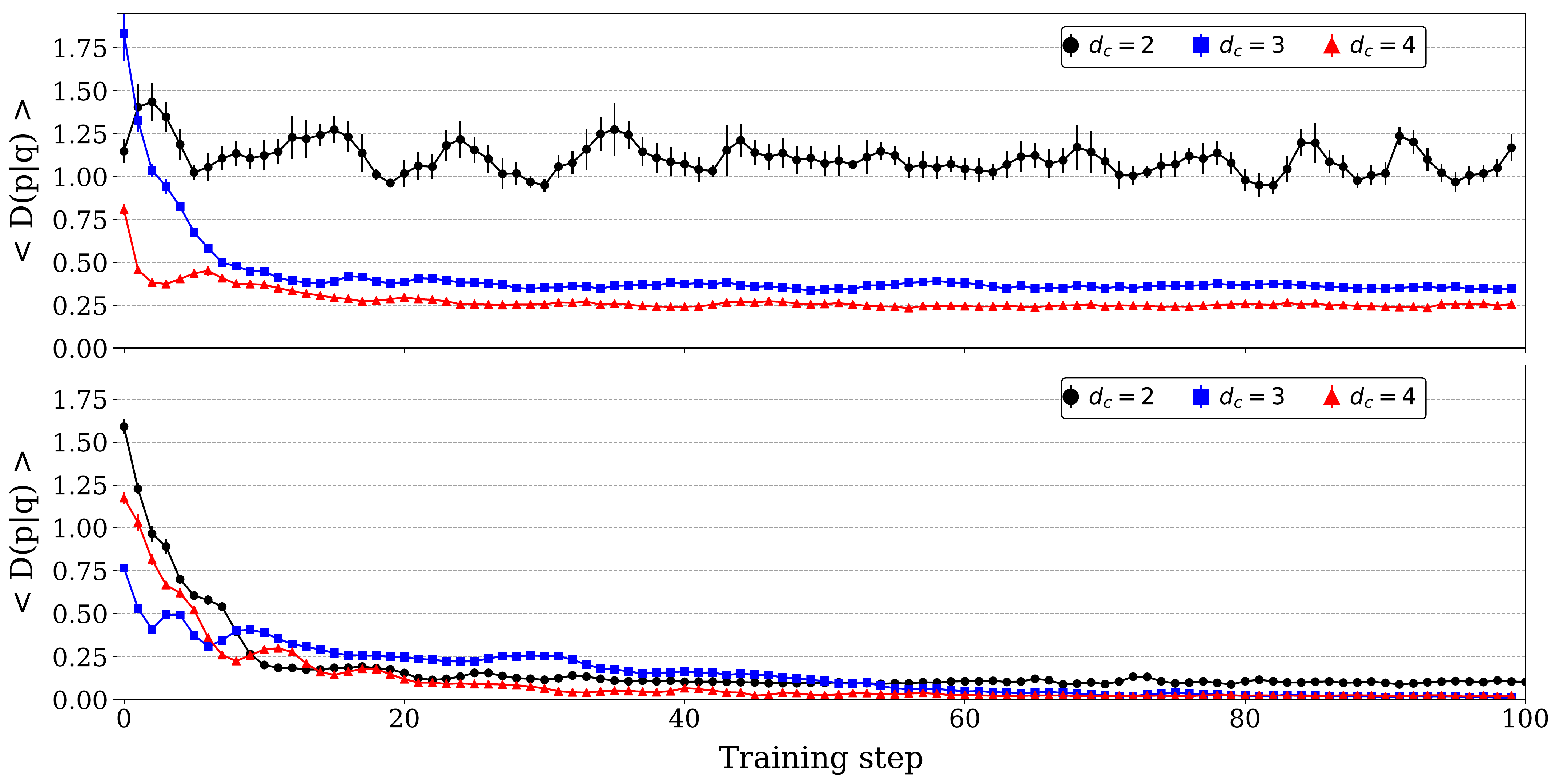}
\caption{KL divergence as a function of training step using $N_{shots}=1024$ during training. Top (bottom) panel corresponds to $L = 1 (2)$ entangling layers.}
\label{fig:KL_sim_comparision}
\end{figure}

In~\Cref{fig:KL_sim_comparision}, training reduces the value of $\langle \KL{p}{q} \rangle$ for the $d_C=3,4,L=1$ circuits, while $\langle \KL{p}{q} \rangle$ of the $d_C=2, L=1$ circuit fluctuates about a quasi-steady mean value $\sim 1.1$. With qubits being entangled pairwise, this ansatz generates a state manifold of the tensor product of two Bell states, up to local rotations. This tensor product structure lacks the complexity to fully learn and describe all of the BAS(2,2) states.  The \Fscore score supports this claim. In~\Cref{sec:appendix_2} we provide additional results for training with smaller learning rates.  

\begin{figure}[htbp]
\centering
\includegraphics[width=\linewidth]{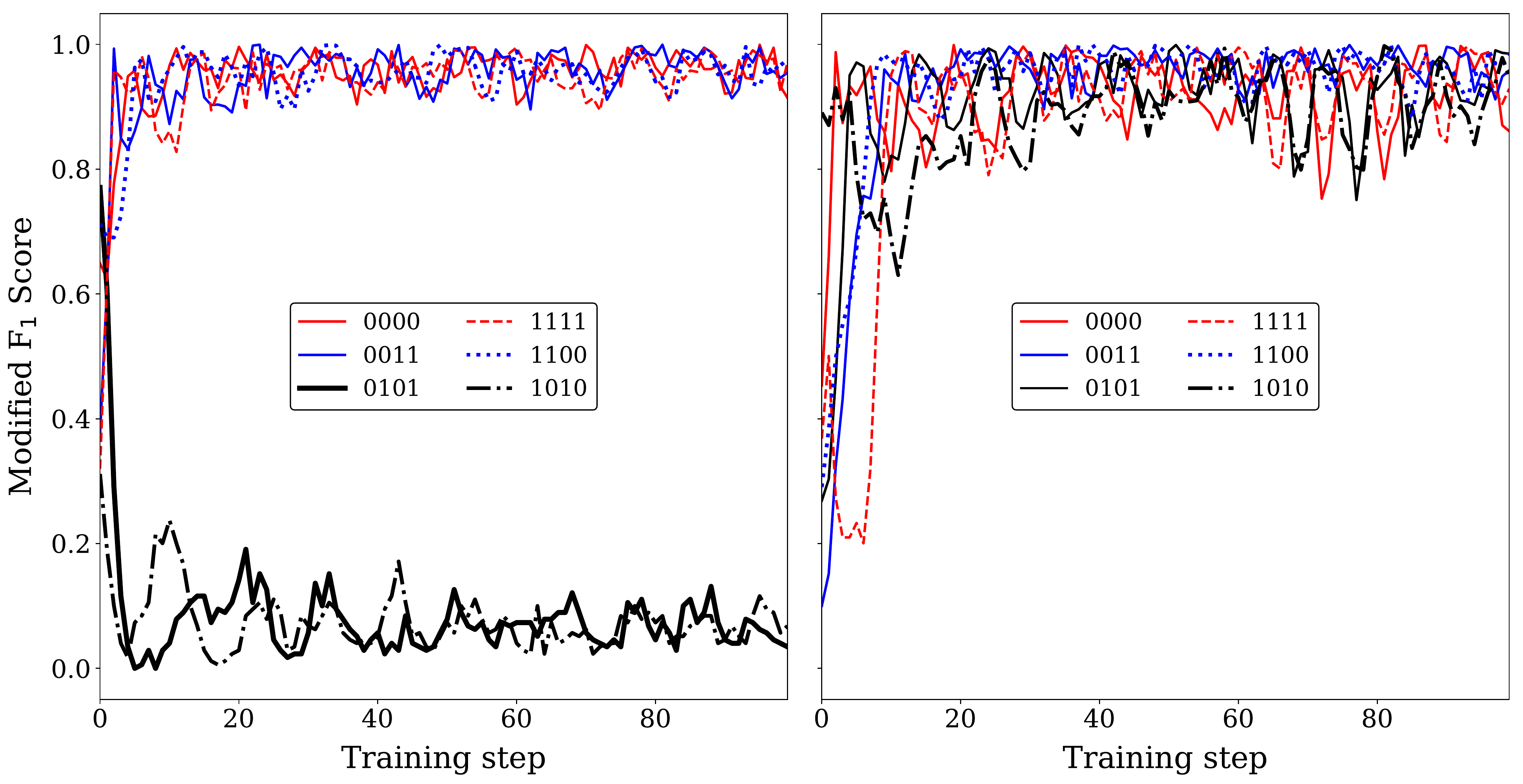}
\caption{ The \Fscore score for each of the $6$ BAS(2,2) states sampled with $N_{shots} = 2048$ at each training step: (Left) ($d_C=2, L=1$) circuit,  (Right) ($d_C=2,L=2$) circuit.}
\label{fig:F1score_sim}
\end{figure}

In~\Cref{fig:F1score_sim}, the individual \Fscore score for each BAS(2,2) state is plotted as a function of training step.  For the $(d_C=2, L=1)$ circuit, it is clear that the QCBM never learns the states $|1010\rangle$ or $|0101\rangle$.  

\begin{figure}[h!]
\centering
\includegraphics[width=\linewidth]{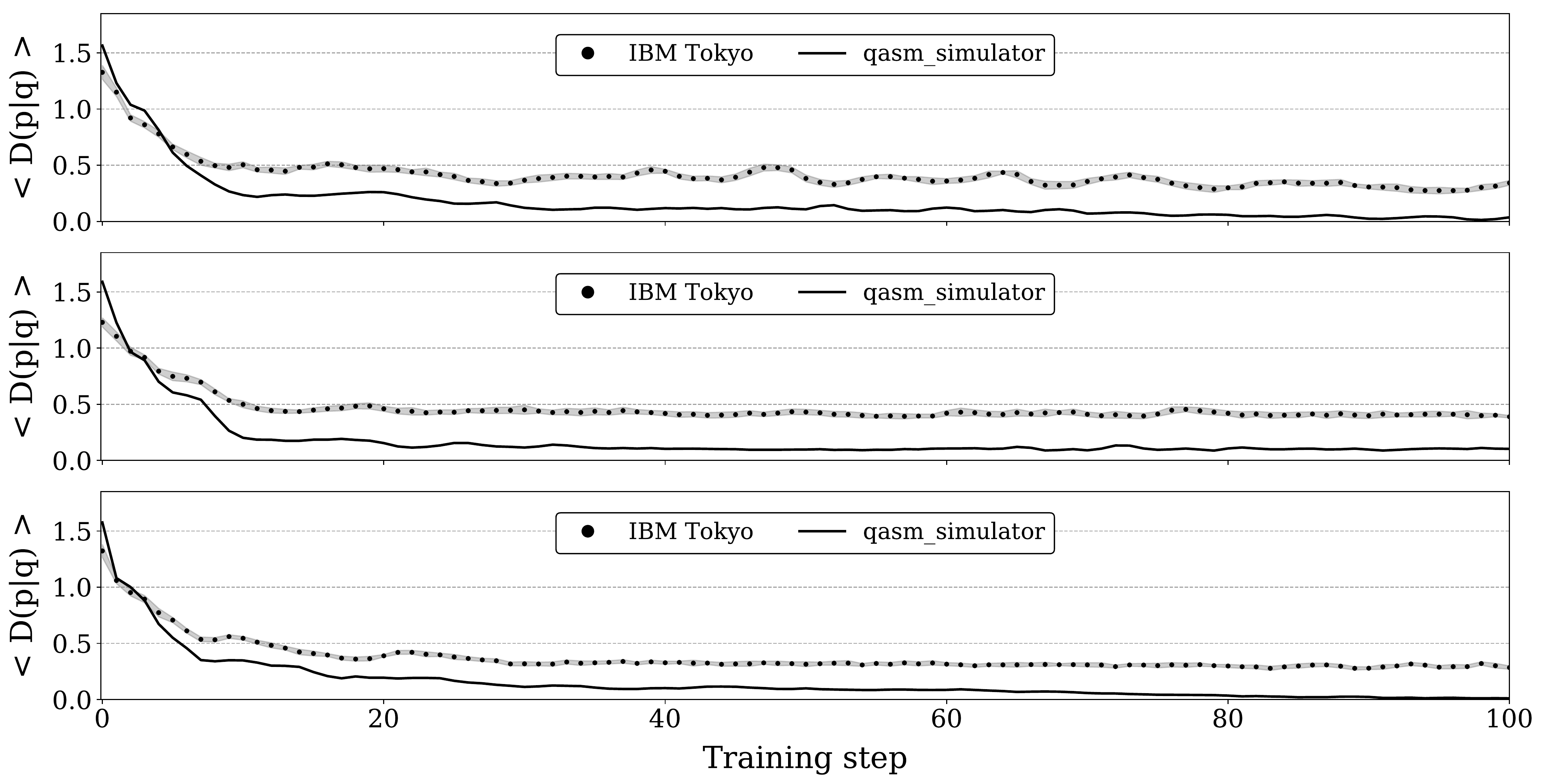}
\caption{Comparison of $\langle \KL{p}{q}\rangle$ for $10$ circuit evaluations of the $d_C=2,L=2$ circuit ansatz deployed on noiseless qubits (black, solid) and noisy qubits (black, circles). (Top) Trained with $N_{shots}=512$,  (Middle) $N_{shots}=1024$, and (Bottom) $N_{shots}=2048$. The standard deviation of $\langle \KL{p}{q}\rangle$ is shown by the grey shaded regions.}
\label{fig:VC_sim_TOKYO}
\end{figure}

\begin{figure}[htbp]
\centering
\includegraphics[width=\linewidth]{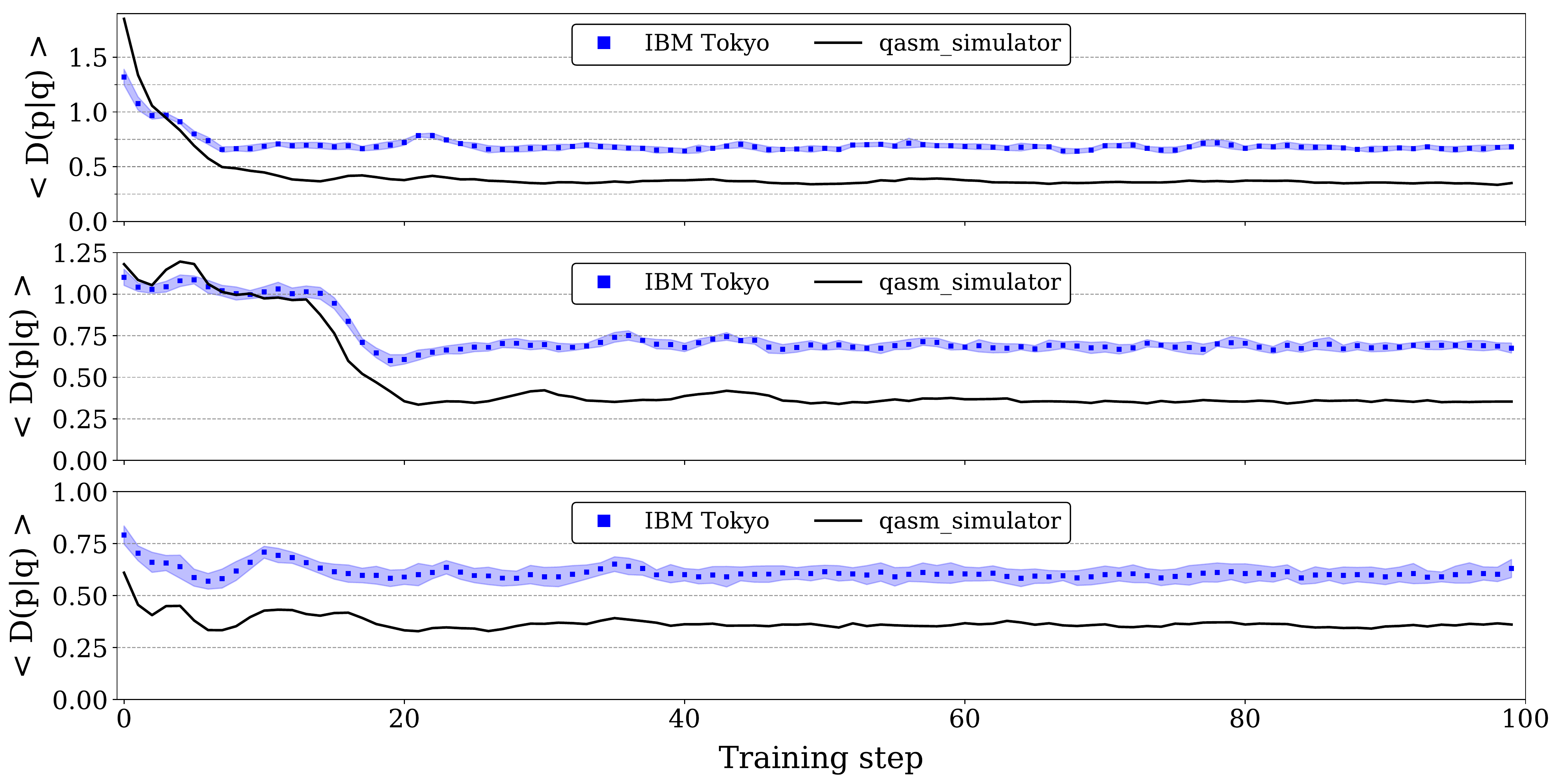}
\caption{Comparison of $\langle \KL{p}{q}\rangle$ for $10$ circuit evaluations of the $d_C=3,L=1$ circuit ansatz deployed on noiseless qubits (black) and noisy qubits (blue, squares). (Top) Trained with $N_{shots}=512$,  (Middle) $N_{shots}=1024$, and (Bottom) $N_{shots}=2048$. The standard deviation of $\langle \KL{p}{q}\rangle$ is shown by the blue shaded regions.}
\label{fig:CL_sim_TOKYO}
\end{figure}

\begin{figure}[htbp]
\centering
\includegraphics[width=\linewidth]{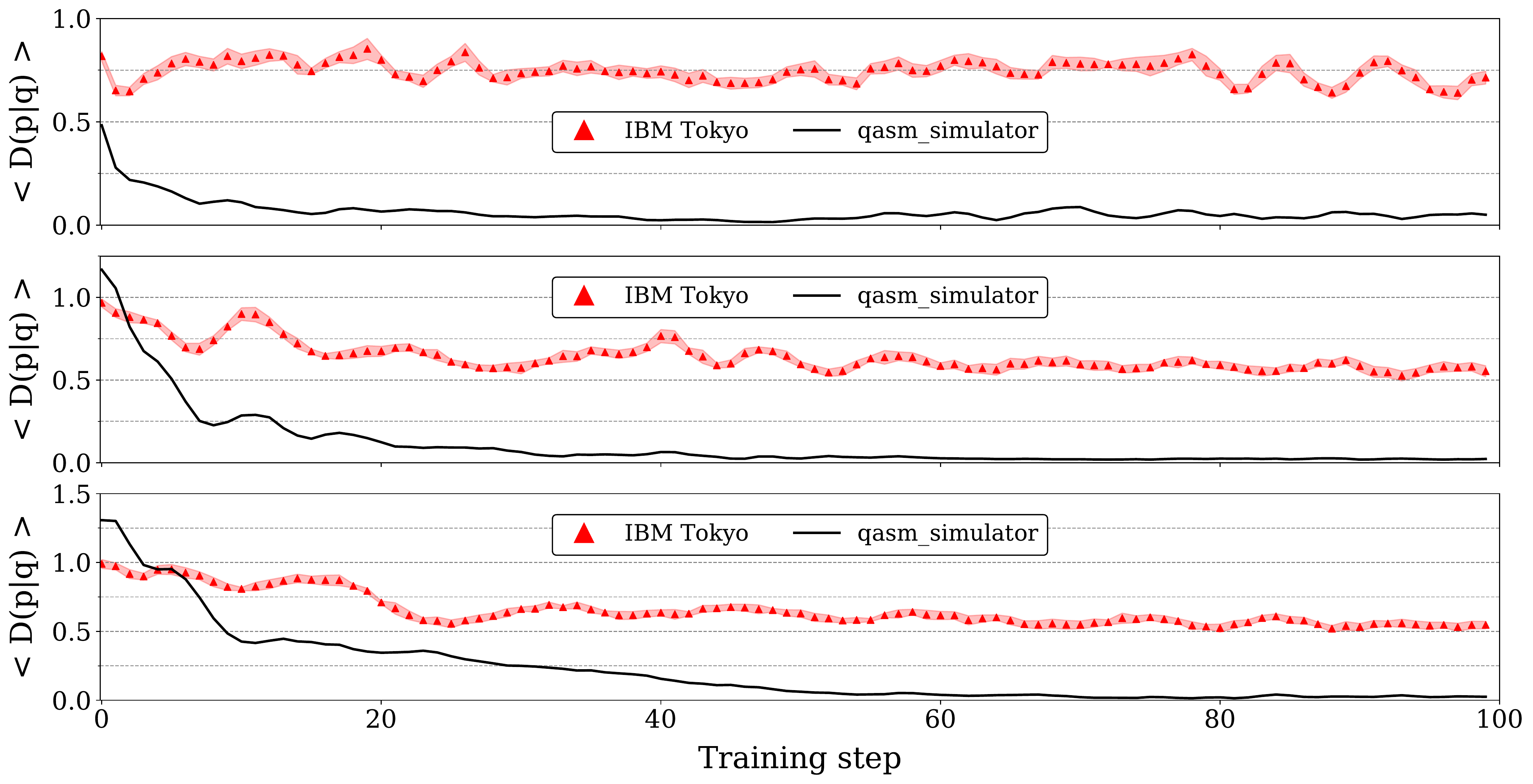}
\caption{Comparison of $\langle \KL{p}{q}\rangle$ for $10$ circuit evaluations of the $d_C=4,L=2$ circuit ansatz deployed on noiseless qubits (black) and noisy qubits (red, triangles). (Top) Trained with $N_{shots}=512$,  (Middle) $N_{shots}=1024$, and (Bottom) $N_{shots}=2048$. The standard deviation of $\langle \KL{p}{q}\rangle$ is shown by the red shaded regions.}
\label{fig:NN_sim_TOKYO}
\end{figure}

We deploy the circuits with trained noiseless parameters on the Tokyo chip to evaluate circuit performance in the presence of noise.  While we leave more detailed discussion about circuit optimization in the presence of noise to~\Cref{sec:discussion}, we show several examples here of how the behavior of $\langle \KL{p}{q} \rangle$ is affected by the addition of noise.  Many circuits show a general offset for $\langle \KL{p}{q} \rangle$, but the behavior on noisy qubits can be substantially different from simulation.  When the QCBM is actively learning ($<30$ training steps) parameter updates which result in large fluctuations on noisless qubits (see~\Cref{fig:VC_sim_TOKYO}) will only result in small changes in $\langle \KL{p}{q} \rangle$ on noisy qubits. When the QCBM training has converged ($>60$ training steps) $\langle \KL{p}{q} \rangle$ reaches a quasi-stationary value for most circuits (c.f.~\Cref{fig:KL_sim_comparision}). 
When deployed on hardware, noise can degrade the efficacy of training on noiseless qubits (see~\Cref{fig:NN_sim_TOKYO}, top).  In contrast, the $d_C=2, L=2$, or $d_C=3, L=1$ circuits reach a quasi-stationary value of $\langle \KL{p}{q} \rangle$ (see~\Cref{fig:VC_sim_TOKYO,fig:CL_sim_TOKYO}) that is lower than the starting value. 

\begin{table}[htbp]
\setlength{\tabcolsep}{8pt}
\centering
\caption{$\min(\langle \KL{p}{q}\rangle)$ circuits evaluated on IBM Tokyo.  Mean calculated over $10$ independent metric evaluations.}
\begin{tabular}{l c c c c}
\hline
L & $N_{shots}$ & $d_C=2$ & $d_C=3$ & $d_C=4$\\
\hline
 1 & 512 & $0.91\pm 0.01$ & $0.64\pm 0.01$ & $0.59\pm 0.02$\\
 1 & 1024 & $0.81\pm 0.02$ & $0.60\pm 0.02$ & $0.54\pm 0.01$\\
 1 & 2048 & $0.86\pm 0.01$ & $0.57\pm 0.02$ & $0.58\pm 0.01$\\
\hline
\end{tabular}
\label{tab:Tokyo_L1_values}
\end{table}

\begin{table}[htbp]
\setlength{\tabcolsep}{8pt}
\centering
\caption{$\min(\langle \KL{p}{q}\rangle)$ circuits evaluated on IBM Tokyo.}
\begin{tabular}{l c c c c}
\hline
L & $N_{shots}$ & $d_C=2$ & $d_C=3$ & $d_C=4$\\
\hline
 2 & 512 & $0.27\pm 0.02$ & $0.48\pm 0.02$ & $0.64\pm 0.02$\\
 2 & 1024 & $0.39\pm 0.01$ & $0.39\pm 0.01$ & $0.53\pm 0.01$\\
 2 & 2048 & $0.28\pm 0.02$ & $0.59\pm 0.01$ & $0.52\pm 0.01$\\
\hline
\end{tabular}
\label{tab:Tokyo_L2_values}
\end{table}

In~\Cref{tab:Tokyo_L1_values,tab:Tokyo_L2_values} we report the best metric values for each $d_C, L$ and $N_{shots}$ value.  The smallest KL value 
was found with the $d_C=2, L=2$ circuit.
When deployed on hardware, increasing the number of rotational parameters improves performance for the $d_C=2,3$ circuits, but not for $d_C=4$ circuits. 

%% file: experiment2_learning.tex
The experiments described in~\Cref{sec:experiment1} explored how closely the value $\KL{p}{q}$ would follow the noiseless learning when measured with noisy qubits.  In this section, we investigate how well QCBM circuits can be trained with a finite number of steps utilizing noisy qubits. The experiments in this section allow us to explore hardware training within the rotational parameter space. 

The goal of these tests is to determine if training a circuit ansatz with noisy qubits can improve the KL metric.  In \Cref{tab:Tokyo_L2_values}, the ($d_C=2,L=2$) circuit reached a minimum value of $0.27(1)$ using theta values trained only with noiseless qubits.  In this section the circuit initialization is chosen at equally spaced intervals from the first $60$ training steps of each of the curves shown in~\Cref{fig:VC_sim_TOKYO}.  This initializes the circuit with: completely random set of parameters ($S=0$), parameters that have undergone some optimization with Adam ($S=10,20,30$), or parameters that have mostly converged to a localized set of values ($S=40,50,60$).  We only train the $(d_C=2, L=2)$ circuit, which was able to reach the lowest value of $\langle \KL{p}{q} \rangle$ with pre-trained parameters (see \Cref{tab:Tokyo_L2_values}). 

As in \Cref{sec:noiseless_training}, the training is done with $3$ shot sizes  $N_{shots}=(512,1024,2048)$ but we evaluate KL metric with $N_{shots}=2048$.  The arithmetic mean value of \KL{p}{q} is calculated from $10$ circuit evaluations at every training step.  We report the following values:  the initial mean value $\langle \dots \rangle_i$, the final value after training $\langle \dots \rangle_f$ and the minimum KL value observed over training.

\begin{table}[htbp]
\setlength{\tabcolsep}{10pt}
\centering
\caption{Trained on IBM Tokyo ($N_{shots} = 512$).  Mean value calculated over $10$ independent metric evaluations.}
\begin{tabular}{l c c c}
\hline
S &  $\langle \KL{p}{q} \rangle_i$ &  $\langle \KL{p}{q} \rangle_f$  &  $\min{\langle\KL{p}{q} \rangle}$  \\
\hline
0 & $1.29\pm 0.05$ & $0.39\pm 0.02$ & $0.39 \pm 0.02$ \\
10 & $0.48\pm 0.02$ & $0.35\pm 0.02$ & $0.35\pm 0.02$ \\
20 & $0.46\pm 0.02$ & $0.34\pm 0.02$ & $0.34\pm 0.02$ \\
30 & $0.32\pm 0.02$ & $0.34\pm 0.01$ & $0.32\pm 0.02$ \\
40 & $0.41\pm 0.02$ & $0.30\pm 0.02$ & $0.29\pm 0.02$ \\
50 & $0.36\pm 0.02$ & $0.30\pm 0.03$ & $0.30\pm 0.03$ \\
60 & $0.36\pm 0.02$ & $0.31\pm 0.02$ & $0.30\pm 0.01$ \\
70 & $0.32\pm 0.02$ & $0.30\pm 0.01$ & $0.29\pm 0.01$ \\
80 & $0.29\pm 0.02$ & $0.32\pm 0.03$ & $0.29\pm 0.02$ \\
\hline
\end{tabular}
\label{tab:Tokyo_2CNOT_512_learning}
\end{table}

\begin{table}[htbp]
\setlength{\tabcolsep}{10pt}
\centering
\caption{Trained on IBM Tokyo ($N_{shots}=1024$).  Mean value calculated over $10$ independent metric evaluations.}
\begin{tabular}{c c c c}
\hline
S &  $\langle \KL{p}{q} \rangle_i$ &  $\langle \KL{p}{q} \rangle_f$  &  $\min{\langle \KL{p}{q} \rangle}$\\
\hline
0 & $1.28\pm 0.05$ & $0.44\pm 0.02$ & $0.43 \pm 0.02 $\\
10 & $0.48\pm 0.03$ & $0.44\pm 0.02$ & $0.43\pm 0.01$\\
20 & $0.49\pm 0.02$ & $0.46\pm 0.02$ & $0.43\pm 0.02$ \\
30 & $0.42\pm 0.02$ & $0.44\pm 0.02$ & $0.42\pm 0.02$ \\
40 & $0.45\pm 0.02$ & $0.50\pm 0.02$ & $0.45\pm 0.02$ \\
50 & $0.44\pm 0.01$ & $0.47\pm 0.02$ & $0.44\pm 0.01$ \\
60 & $0.41\pm 0.04$ & $0.44\pm 0.03$ & $0.41\pm 0.04$ \\
70 & $0.47\pm 0.02$ & $0.46\pm 0.03$ & $0.45 \pm 0.03$ \\
80 & $0.45\pm 0.03$ & $0.46\pm 0.02$ & $0.42 \pm 0.01$ \\
\hline
\end{tabular}
\label{tab:Tokyo_2CNOT_1024_learning}
\end{table}

For completely random initial parameters ($S=0,10$), training with noisy qubits was able to reduce $\langle \KL{p}{q} \rangle$.  However, training that began at later points tend to return higher values of $\langle \KL{p}{q} \rangle$ or show minimal improvement of $\langle \KL{p}{q} \rangle$ after $10$ training steps.  We discuss the effects of noise, shot size and stochastic gradient learning on circuit training in~\Cref{sec:discussion}.
\begin{table}[htbp]
\setlength{\tabcolsep}{10pt}
\centering
\caption{Trained on IBM Tokyo ($N_{shots}=2048$).  Mean value calculated over $10$ independent metric evaluations.}
\begin{tabular}{c c c c}
\hline
S &  $\langle \KL{p}{q} \rangle_i$ &  $\langle \KL{p}{q} \rangle_f$  &  $\min{\langle \KL{p}{q} \rangle}$\\
\hline
0 & $1.30\pm 0.06$ & $0.34\pm 0.02$ & $0.34\pm 0.02$ \\
10 & $0.58\pm 0.01$ & $0.37\pm 0.02$ & $0.37\pm 0.02$ \\
20 & $0.37\pm 0.02$ & $0.35\pm 0.02$ & $0.30\pm 0.02$ \\
30 & $0.30\pm 0.02$ & $0.36\pm 0.2$ & $0.30\pm 0.02$ \\
40 & $0.33\pm 0.01$ & $0.35\pm 0.01$ & $0.33\pm 0.01$ \\
50 & $0.38\pm 0.02$ & $0.43\pm 0.03$ & $0.38\pm 0.02$ \\
60 & $0.29\pm 0.02$ & $0.34\pm 0.02$ & $0.29\pm 0.02$ \\
70 & $0.30\pm 0.03$ & $0.29\pm 0.02$ & $0.29\pm 0.02$ \\
80 & $0.29\pm 0.02$ & $0.30\pm 0.03$ & $0.29\pm 0.02$ \\
\hline
\end{tabular}
\label{tab:Tokyo_2CNOT_2048_learning}
\end{table}

%% file: discussion.tex
Effective classical machine learning relies on proper tuning of hyper-parameters and avoiding over-fitting.  By limiting the number of training steps and rotational parameters our models try to fit, we believe that we have avoided circuit ansatz that are too complex for the dataset.  The hyper-parameters of Adam were optimized using noiseless simulation and good rotational parameters were learned for the circuits in this paper, with the exception of the $d_C =2,L=1$ circuit which we will exclude from discussion in this section.  In this section we will use the Kullback-Leiber
divergence to discuss the qualitative changes in performance due to qubit noise and finite sampling.  

\subsection{Device noise}
When simulated with noiseless qubits, increasing the number of rotational parameters improves the capabilities of the QCBM. The lowest $\langle \KL{p}{q} \rangle \sim 0.01$ values were found for $L=2$, regardless of $d_C$ value.  This same convergence is not seen when circuits are deployed on noisy hardware.  Current quantum devices have many sources of noise including: qubit decoherence, gate infidelity, and measurement errors. In this study we assume the training will be able to compensate for noise in the single qubit gates, and the limited circuit size will mitigate decoherence effects.  In this initial study we have not included any readout error mitigation, and designed entangling layers to reduce the noise from 2 qubit CNOT gates.  With the addition of noise the $d_C=2, L=2$ circuit returned the lowest value $\langle \KL{p}{q} \rangle=0.27 \pm 0.02$ using values pre-trained via noiseless simulation.  Comparable values are found when a circuit was trained on noisy qubits, the lowest value found after training was $\langle \KL{p}{q} \rangle=0.29 \pm 0.01$ (see \Cref{tab:Tokyo_2CNOT_512_learning,tab:Tokyo_2CNOT_1024_learning,tab:Tokyo_2CNOT_2048_learning}). Understanding how training is affected by the loss function space is an active area of research for classical machine learning \cite{li2017visualizing,chaudhari2018stochastic}. We will use this concept to frame our discussion in this section using $\tau_{U}(\tau_{U^{\prime}})$ for to the loss function space of a noiseless (noisy) circuit.

For a circuit with $R$ rotational parameters, the loss function space $\tau$ is defined over the $R$ dimensional set of all possible parameter values.  We will compare the noiseless and noisy qubit performances to draw conclusions about how the addition of noise affects the space $\tau_{U}$ of a single circuit ansatz (c.f.~\Cref{fig:VC_sim_TOKYO,fig:CL_sim_TOKYO,fig:NN_sim_TOKYO}) and rely on several assumptions made without explicit models of these spaces. First, varying the value of $d_C$ modifies the encoded degrees of entanglement. The local and global optimal parameters of circuits with different $d_C,L$ will therefore be quite different.  Also, for circuits with the same values of $d_C, L$ noise will cause the spaces $(\tau_{U}, \tau_{U^{\prime}})$ to differ.  

In the absence of qubit noise the training has largely converged after $\approx 50$ steps of training. With the weight decay implemented in Adam, this implies that the optimizer is taking small steps within a localized region of $\tau_{U}$.  Our first observation is trivial:  just as the optima of $\tau_{U}$ are expected to be different for different $d_C,L$ values; the minimum that Adam converges to in $\tau_{U}$ is not guaranteed to be a minimum in $\tau_{U^{\prime}}$ and using Adam to optimize over $\tau_{U}$ instead may drive the system further from the ideal parameters for $\tau_{U^{\prime}}$. However, small changes in parameters can lead to a good minimum within the space $\tau_{U^{\prime}}$.  Secondly, the stability of $\tau_{U}$, does not necessarily predict the stability of $\tau_{U^{\prime}}$. Small changes in parameters can lead to fluctuations in $\langle \KL{p}{q} \rangle$ or possibly degredation on noisy qubits (c.f.~\Cref{fig:NN_sim_TOKYO}, $N_{shots}=512$).  On the other hand, the convergence in $\tau_{U}$ to an improved value can be seen in $\tau_{U^{\prime}}$ (c.f.~\Cref{fig:VC_sim_TOKYO}, $N_{shots}=1024$); the relative stability of the KL divergence implies that Adam is exploring a region of $\tau_{U}$ which is quasi-stable in $\tau_{U^{\prime}}$. 

Rotational parameters learned during training are dependent on hardware noise and variability.  For all values of $N_{shots}$, training on hardware improved $\langle \KL{p}{q} \rangle $ when the circuit was initialized with a random set of parameters, or pre-trained parameters obtained from a low number of Adam steps ($S<40$).  On the other hand, continued training on hardware after the training in the simulator has already converged yields no improvement in $\langle \KL{p}{q} \rangle $. The hardware-trained parameters overall yielded less of an improvement than trained parameters from the simulator due to the inherent noise of the quantum computer. Therefore, interleaving error mitigation steps with each training step is expected to improve performance of hardware-trained parameters, and this is the subject of a future study.
\subsection{Sampling}
In~\Cref{sec:experiment1} we trained multiple circuits from random initial values using noiseless qubits.  For each circuit, Adam trains a unique QCBM and defines a unique path in a $16(28)$-dimensional space for $L=1(L=2)$ circuits.  Within $100$ training steps the optimizer is able to find local minima, however it is not guaranteed to converge to the global optima (see~\Cref{tab:sim_L2_values}). The noise introduced by smaller $N_{shots}$ values could improve exploration during training. 

Sampling a circuit with a high number of shots can improve the  KL metric evaluation by reducing the probability of erroneously populated states. However reducing the sampling error by increasing $N_{shots}$ alone may not be sufficient to counteract the effects of noise on the overall performance of a given circuit.

%% file: conclusions.tex
As quantum devices become available there is a growing need for a cohesive set of benchmarks quantifying hardware performance.  We have observed that while limited connectivity between qubits and noisy gates are not a significant obstacle to circuit learning, our results show that circuit ansatz design can affect generative modeling performance. 

There are $6$ possible CNOT gates that can be defined between pixels of the BAS(2,2) images, and the $d_C=2,4$ circuits show that the distribution can be modeled by placing CNOT gates between neighboring pairs of pixels.  While larger image sizes require long range correlations, efficient encoding of larger datasets into hardware with fixed qubit connectivity remains an open question (see Appendix \ref{sec:appendix_4}). For the BAS(2,2) dataset, adding more CNOTs to a single qubit in each entangling layer led to minimal increases in performance on noisy qubits. When deployed on hardware, the $d_C=2,L=2$ circuit outperformed all other circuits. 

Using a noise-robust stochastic optimizer allows us to train quantum circuits in the presence of noisy hardware. The provided metrics show the hardware's capability to reproduce desired probability distributions in the presence of both systematic and statistical noise. 
We also observe that measurement shot noise can minimally affect the training of a QCBM.  However, classical effects such as the optimizer getting trapped in local minima are more significant.

Since the hardware is both noisy and has somewhat sparse connectivity, choosing entangling layers with sufficient sparseness to avoid excessive systematic error while still providing enough complexity to reproduce the distribution represents a trade-off that can be explored using the metrics as a guide. Evaluating the metric for a few entangling layer designs gives insight into which entanglement circuits are good at providing the complexity to represent certain distributions with low noise.

Further development of this benchmark will focus on improvements to the noise-resilience of circuit training which will lead to better estimates of the hardware's innate capabilities. Areas of development include: incorporating error mitigation~\cite{kandala_error_2019} into circuit training to counteract the effects of measurement (readout) and gate errors, and exploring other classical optimizers to find the most robust methods for a given hardware device. The benchmark presented in this work is a useful measure of a quantum computer ability to reproduce a discrete probability distribution, and we demonstrated its utility by analyzing the performance of a superconducting quantum computer. While fully noise-robust circuit learning remains an open question, as a benchmark it shows promising avenues for future application and refinement.

%% file: appendix1.tex
\section{Alternate performance metrics}
\label{sec:appendix_1}
 A metric introduced in \cite{benedetti2018generative} called the qBAS22 score can be used to evaluate how well a circuit modeled the BAS(2,2) distribution.  An advantage to using the qBAS22 score is that it remains finite even if a BAS state is absent from the sample distribution.  In this section we report values of the qBAS22 score for reference.

Accurately measuring the qBAS22 score relies on a large number samples drawn from a circuit with low sample size.  In the Appendix of \cite{benedetti2018generative} the sample size needed to evaluate the qBAS22 score is derived for different BAS(n,m) distributions (for BAS(2,2) it is $15$).  

We measure and report the qBAS22 score at each training step for the $6$ circuits introduced in the main text and focus on circuits trained with $N_{shots}=1024$.  At each training step we evaluate a given circuit $11$ times with $N_{shots}=1024$, generating $11$ independent distributions.  After a distribution is obtained from a circuit using $N_{shots}=1024$, we then draw $10,000$ samples of size $15$ (sampling done with replacement).  Then we evaluate the qBAS22 score $11$ times and report the weighted arithmetic mean value of $\langle \qBAS \rangle$.  

\begin{figure}[htbp]
\centering
\includegraphics[width=\linewidth]{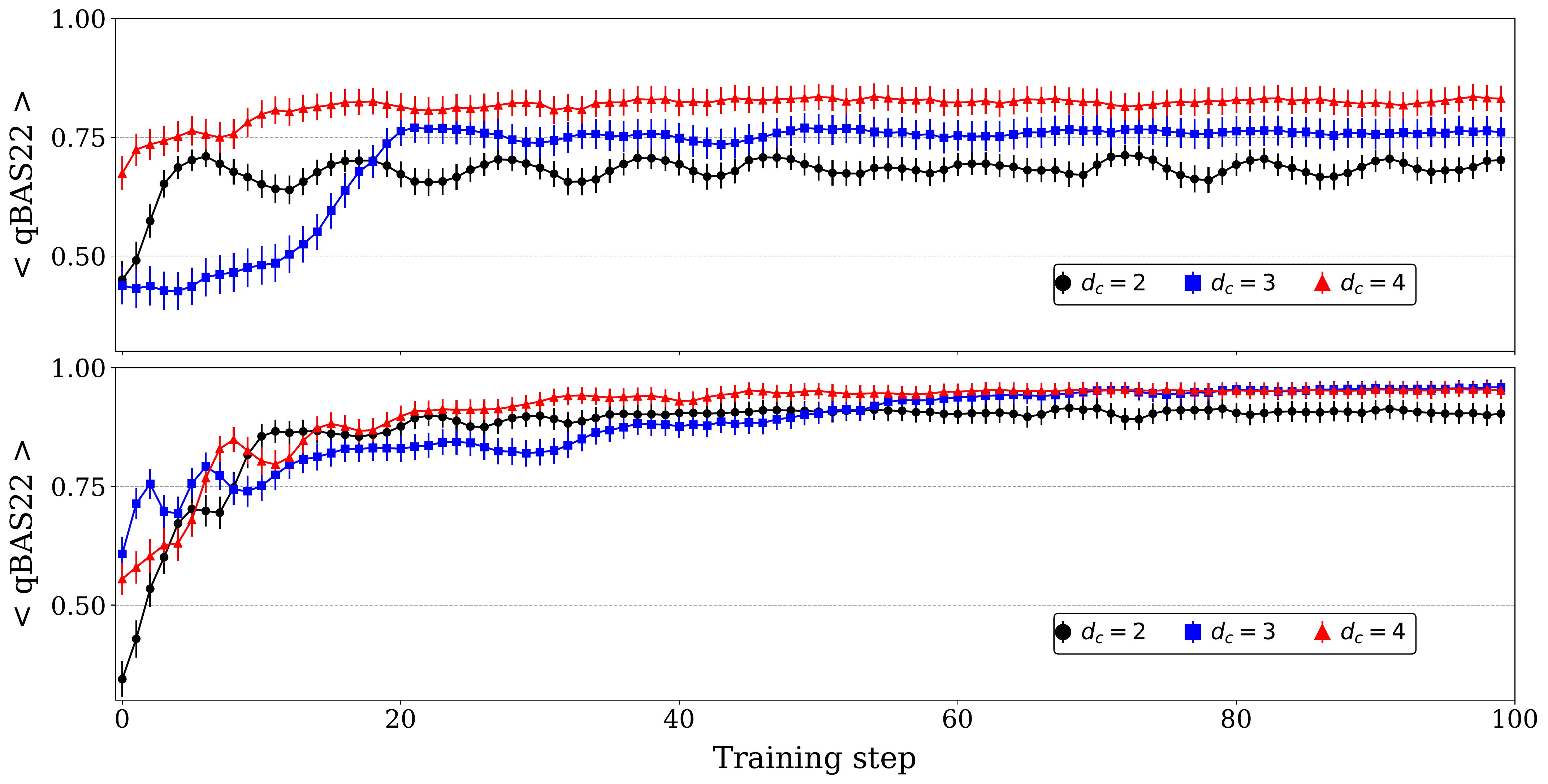}
\caption{The $\langle \qBAS \rangle$ scores evaluated at each training step using $N_{samples}=15$ and $10000$ samples.  The mean is defined by the weighted mean taken over $11$ independent distributions sampled from a circuit with $N_{shots}=1024$. The error bars are defined by the weighted variance.(Top)  For the $L=1$ circuits trained and evaluated on noiseless qubits. (Bottom) For the $L=2$ circuits trained and evaluated on noiseless qubits. }
\label{fig:qBAS_sim_comparision}
\end{figure}

We evaluate this metric for circuits trained with noiseless qubits and for circuits trained on hardware. In Fig.~\ref{fig:qBAS_sim_comparision} we show the qBAS22 score for circuits that are trained with noiseless qubits and the metric is evaluated with noiseless qubits.  The qBAS22 score for the $d_C=2, L=1$ circuit (which doesn't completely model the entire BAS(2,2) dataset) is the lowest performing circuit. Of the $6$ circuits shown in Fig.~\ref{fig:qBAS_sim_comparision} the $L=2, d_C=3,4$ circuits have the highest qBAS22 scores ($0.96\pm0.04$ and $0.95\pm0.4$, respectively).

\begin{figure}[htbp]
\centering
\includegraphics[width=\linewidth]{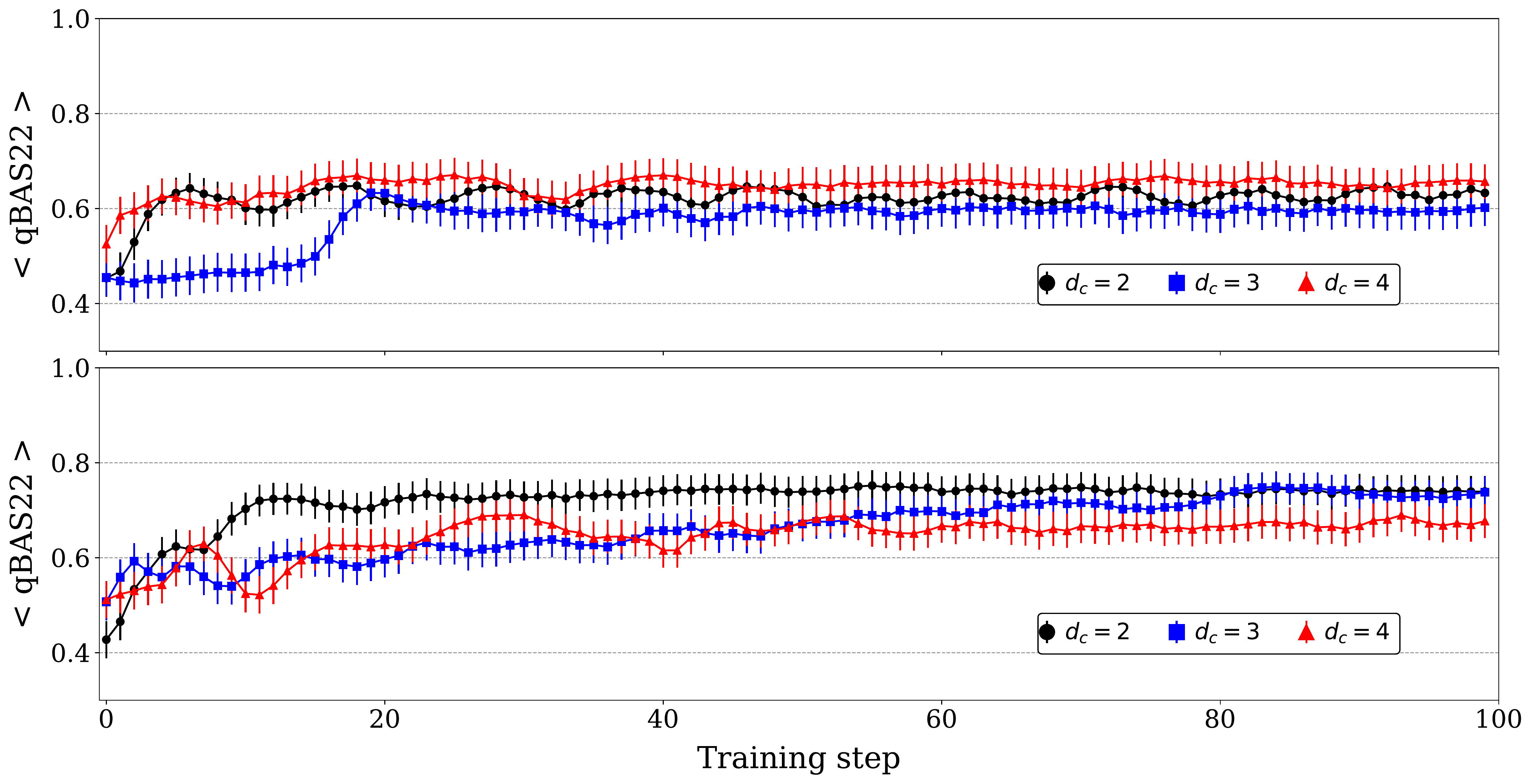}
\caption{The $\langle \qBAS \rangle$ scores evaluated at each training step using $N_{samples}=15$ and $10000$ samples.  The mean is defined by the weighted mean taken over $11$ independent distributions sampled from a circuit with $N_{shots}=1024$. The error bars are defined by the weighted variance.(Top)  For the $L=1$ circuits trained on noiseless qubits and evaluated on IBM Tokyo. (Bottom) For the $L=2$ circuits trained on noiseless qubits and evaluated on IBM Tokyo. }
\label{fig:qBAS_Tokyo_comparision}
\end{figure}

However the device noise strongly affects the $d_c=3,4$ circuits.  In ~\Cref{fig:qBAS_Tokyo_comparision} we present the qBAS22 scores for circuits trained wiht noiseless qubits, but evaluate the metric on IBM Tokyo.  We see that for $L=1$ circuits the $d_C=2$ circuit perform comparably to the $d_C=3,4$ circuits, even though this circuit is known to only fit $4$ out of the $6$ BAS(2,2) states. When the circuit size is increased to $L=2$, the $d_c=2,3$ circuits have comparable performance after $100$ steps of training ($0.75\pm0.04$ and $0.75\pm0.04$, respectively), out-performing the $d_c=4$ circuit ($0.69\pm0.04$).  Similar behavior is seen in the KL metric reported in the main text (c.f. \Cref{tab:Tokyo_L2_values}).

In ~\Cref{tab:qBAS_training_batch1024} we present the qBAS22 score evaluated on IBM Tokyo for the ($d_C=2,L=2$) circuit trained on hardware. As in~\Cref{sec:experiment2} the circuits are pre-trained using noiseless simulation for a fixed number of steps, then deployed on IBM Tokyo hardware to execute $10$ steps of Adam training.  The best performance of a circuit trained on hardware for $10$ steps of Adam was $\langle \qBAS \rangle = 0.74\pm0.03$.  
\begin{table}[htbp]
\setlength{\tabcolsep}{10pt}
\centering
\caption{The mean is defined by the weighted mean taken over $11$ independent distributions sampled from a circuit with $N_{shots}=1024$. Circuit trained on IBM Tokyo with $N_{shots} = 1024$ and metric evaluated on IBM Tokyo.}
\begin{tabular}{c c c c}
\hline
S &  $\langle \qBAS \rangle_i$ &  $\langle \qBAS \rangle_f$  &  $\max{\langle \qBAS \rangle}$  \\
\hline
0  & $0.42\pm0.04$ & $0.74\pm0.03$ & $0.74\pm0.03$\\
10 & $0.71\pm0.03$ & $0.72\pm0.03$ & $0.74\pm0.03$\\
20 & $0.70\pm0.03$ & $0.71\pm0.03$ & $0.72\pm0.03$\\
30 & $0.74\pm0.03$ & $0.73\pm0.03$ & $0.74\pm0.03$\\
40 & $0.73\pm0.03$ & $0.69\pm0.04$ & $0.73\pm0.03$\\
50 & $0.73\pm0.03$ & $0.71\pm0.03$ & $0.73\pm0.03$ \\
60 & $0.74\pm0.03$ & $0.72\pm0.03$ & $0.74\pm0.03$ \\
70 & $0.72\pm0.03$ & $0.71\pm0.03$ & $0.72\pm0.03$ \\
80 & $0.72\pm0.03$ & $0.71\pm0.03$ & $0.73\pm0.03$ \\
\hline
\end{tabular}
\label{tab:qBAS_training_batch1024}
\end{table}

The qBAS22 metric and the KL metric give a measure of the global performance of a circuit but there is also a need for local metrics.  We adapt the \Fscore score \cite{goodfellow2016deep}, and apply it to the individual BAS(n,m) states to define a metric that measures how well a circuit learns each state and can be applied to uniform or non-uniform discrete distributions. However, it requires that the user specify the exact form of the target distribution.  For benchmarking tasks where the performance is measured with regards to a known distribution this is not a problem, but it may limit the usability of the \Fscore score metric for future applications.  

The \Fscore score relies on the precision and true positive rate of a model and in our metric these quantities are defined with respect to the uniform BAS(2,2) distribution ($p_i = 1/6$ if $|x_i\rangle$ is a BAS(2,2) state).  Device noise (such as readout errors) leads to a number of incorrectly measured states, but in our initial approximation, for each state $|x_i \rangle$ of the BAS dataset we define the number of true positives as $\mathrm{TP}(x_i)=q(x_i)$, i.e. the sampled probability of the state $|x_i\rangle$. We define the number of false positives (FP) and false negatives (FN) using the difference $\Delta = |q(x_i)-p(x_i)|$. If $q(x_i)>p(x_i)$ then $\mathrm{FP}(x_i) = \Delta$ and $\mathrm{FN}(x_i) =0$; if $q(x_i)<p(x_i)$ then $\mathrm{FN}(x_i) = \Delta$ and $\mathrm{FP}(x_i)=0$.  For each state $x_i$ we use the true positive rate
\begin{equation}
    \mathrm{TPR}(x_i) = \frac{\mathrm{TP}(x_i)}{[\mathrm{TP}(x_i)+\mathrm{FN}(x_i)]},
\end{equation}
and the precision
\begin{equation}
    \mathrm{P}(x_i)=\frac{\mathrm{TP}(x_i)}{[\mathrm{TP}(x_i)+\mathrm{FP}(x_i)]}.
\end{equation}
The balanced \Fscore score is the harmonic mean of the precision and true positive rate,
 \begin{equation}
     \Fscore(x_i) = 2 \left(\frac{\mathrm{P}(x_i) \times \mathrm{TPR}(x_i)}{\mathrm{P}(x_i)+\mathrm{TPR}(x_i)}\right).
 \end{equation}
 
\section{Alternate learning rates}
\label{sec:appendix_2}
In this section we highlight the specific case of the ($d_C=2, L=1$) circuit.  In ~\Cref{fig:KL_sim_comparision} the KL value oscillated around $\KL{p}{q}\sim 1.1$ and in~\Cref{sec:experiment1,sec:discussion} we argue that this behavior is due to the circuit being overly simplistic and not from a too-large learning rate.  

\begin{figure}[htbp]
\centering
\includegraphics[width=\linewidth]{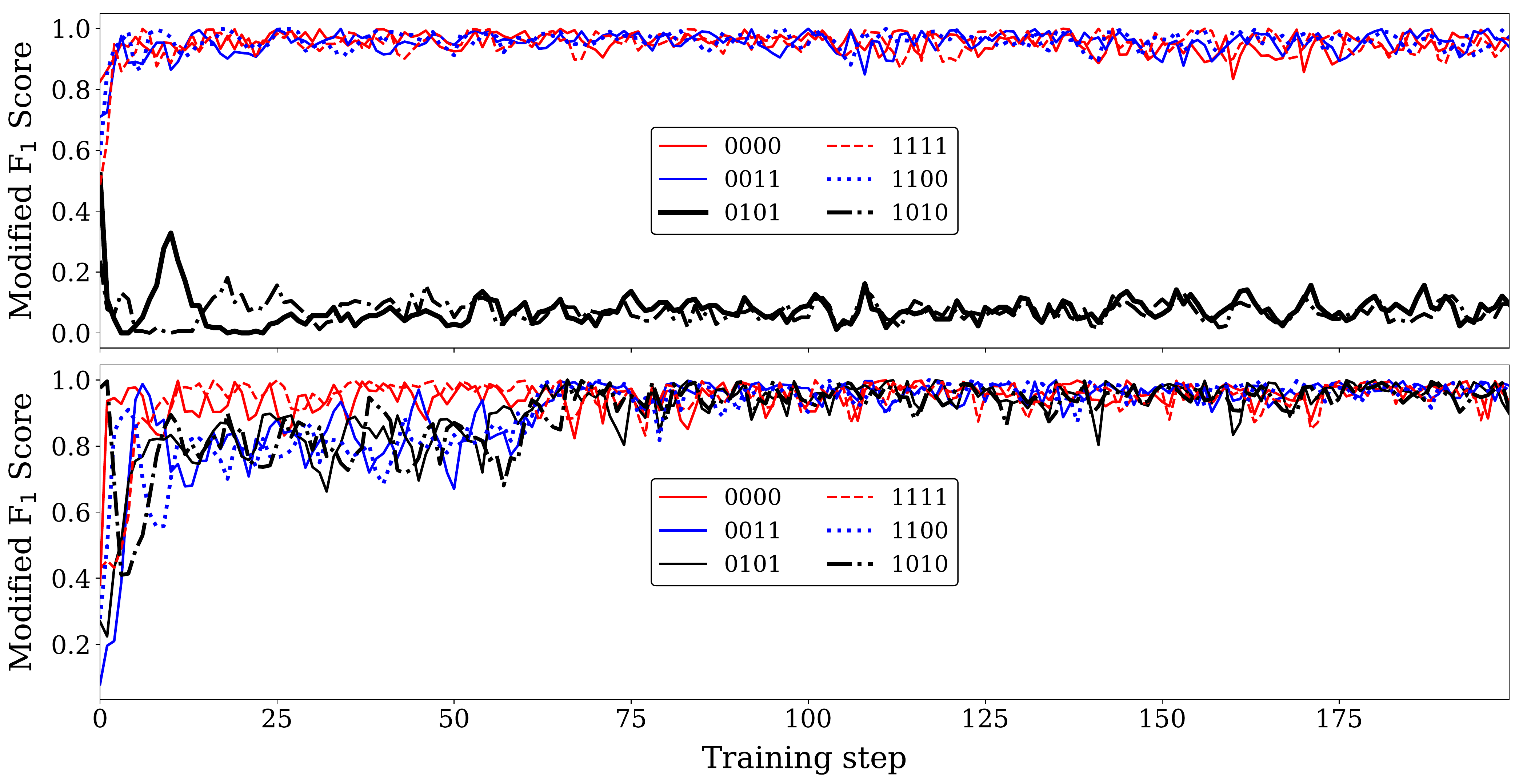}
\caption{\Fscore of a circuit trained on a noiseless simulator with: $N_{shots}=1024$, $200$ steps of ADAM, $\alpha = 0.05$, and sampled with $N_{shots} = 2048$. (Top) $d_C=2, L=1$, (Bottom) $d_C=2, L=2$.}
\label{fig:F1_alpha_depth2}
\end{figure}

\begin{figure}[htbp]
\centering
\includegraphics[width=\linewidth]{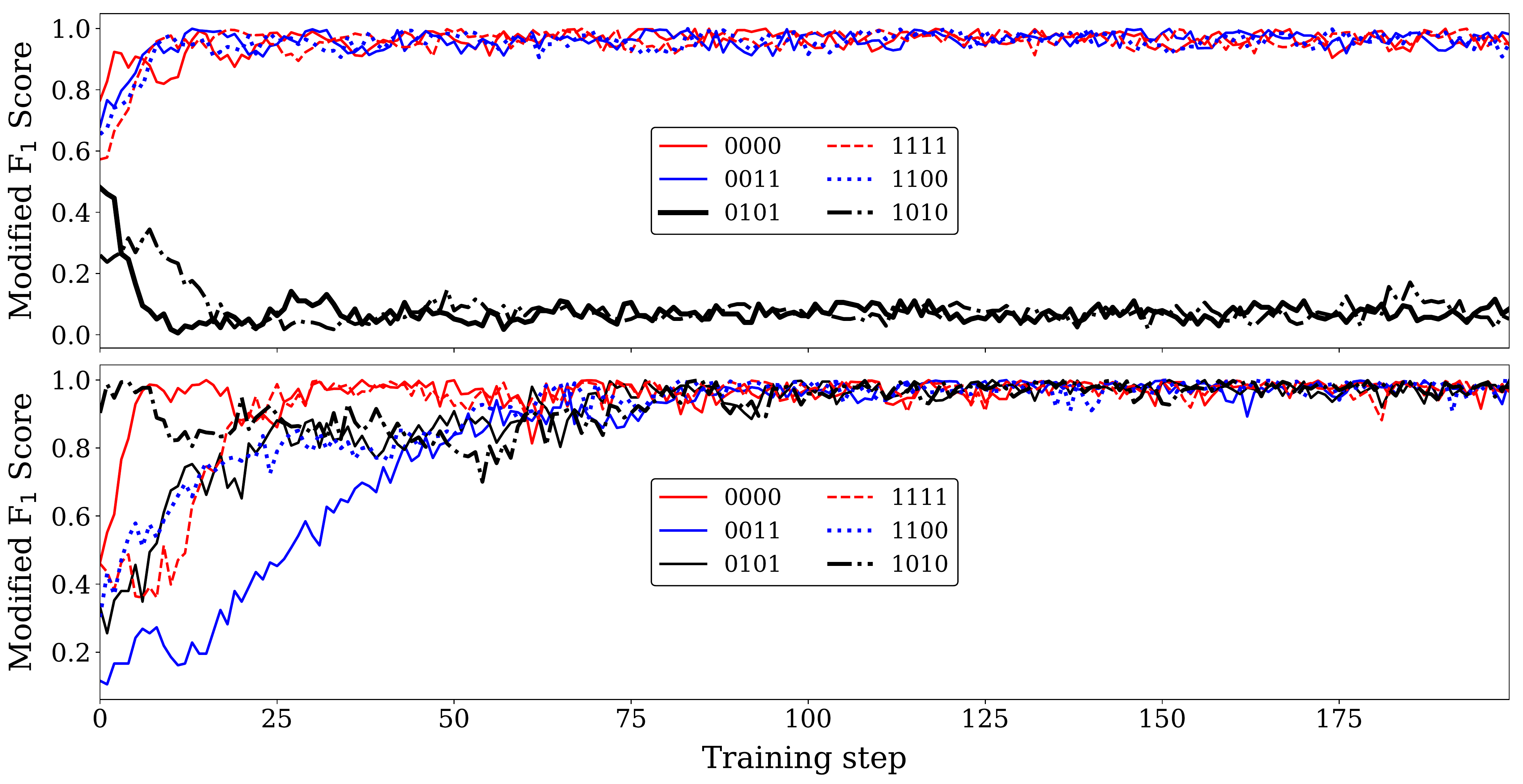}
\caption{\Fscore of a circuit trained on a noiseless simulator with: $N_{shots}=1024$, $200$ steps of ADAM, $\alpha = 0.3$, and sampled with $N_{shots} = 2048$. (Top) $d_C=2, L=1$,  (Bottom) $d_C=2, L=2$.}
\label{fig:F1_alpha_depth1}
\end{figure}

To prove this we re-trained the ($d_C=2,L=1$) circuit with $N_{shots}=1024$ and different learning rates $\alpha = \lbrace 0.05, 0.3\rbrace$.  The circuits were initialized with a random set of angles and trained for $200$ steps of ADAM.  Using the \Fscore score, we see that lowering the learning rate ($\alpha=0.05$) shows no significant improvement (see~\Cref{fig:F1_alpha_depth2}), the ($d_C=2, L=1$) circuit still fails to learn the states $|1010 \rangle$ and $|0101 \rangle$.  In contrast, ($d_C=2, L=2$) circuit is able to learn all $6$ BAS states, even with a higher learning rate ($\alpha=0.3$)    (see~\Cref{fig:F1_alpha_depth1}).
\begin{figure}[htbp]
\centering
\includegraphics[width=\linewidth]{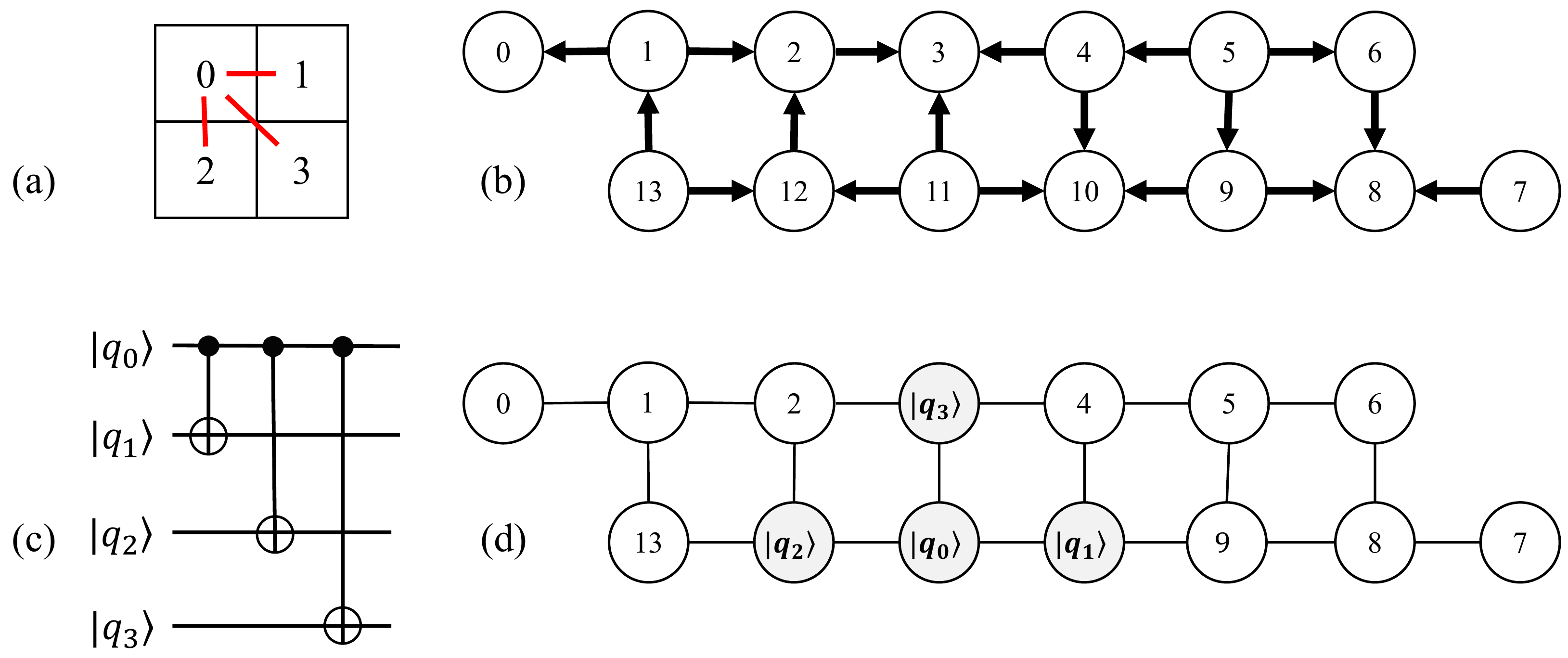}
\caption{(a) The pixels of a BAS(2,2) image with the edges of the Chow-Liu tree defined from the mutual information (red).  (b) The connectivity graph of IBM's Melbourne chip \cite{IBM_Melbourne}.  (c) The $d_C=3$ entangling layer defined using the Chow-Liu tree in (a).  (d) The $d_C=3$ layer embedded into IBM Melbourne.}
\label{fig:locality}
\end{figure}
\section{Connectivity, correlation locality, and hardware embedding}
\label{sec:appendix_4}
We define local or non-local connections with respect to the image pixels of the BAS(2,2) dataset.  There are $6$ possible pairs that can be formed from the four pixels of each image ($4$ local, $2$ non-local).  The nearest neighbor pairs of pixels $[(0,1),(0,2),(1,3),(2,3)]$ form the local connections, while the remaining pairs $[(0,3),(1,2)]$ are non-local.  

If the hardware supports all-to-all connectivity then all local and non-local connections can be mapped to CNOT gates and implemented in a single QCBM.  With limited qubit connectivity, it is possible to embed to non-local connections into hardware but often at the cost of removing local connections.  The $d_C=2,4$ layers construct QCBMs with $4$ local connections and $0$ non-local connections, whereas the $d_C=3$ layers construct QCBMs with $1$ non-local and $2$ local connections.  In \Cref{fig:locality} we show the construction and hardware embedding of a $d_C=3$ entangling layer from the edges of a Chow-Liu tree rooted at pixel $0$.  Understanding the trade-offs between local or non-local connections will be necessary to construct QCBMs that can model larger images or more complicated distributions.  